\documentclass[aps,prd,10pt,tightenlines,notitlepage,showpacs,nofootinbib,superscriptaddress,twocolumn,eqsecnum,raggedbottom,preprintnumbers,longbibliography]{revtex4-1}
\usepackage[export]{adjustbox}
\usepackage{amsfonts}
\usepackage{amsmath}
\usepackage{amssymb}
\usepackage{array}
\usepackage{bbold}
\usepackage{bm}
\usepackage{slashed}
\usepackage{booktabs}
\usepackage[dvipsnames,usenames]{color}
\usepackage{enumitem}
\usepackage{float}
\usepackage{graphicx}
\usepackage{subfigure}
\usepackage[utf8]{inputenc}
\usepackage{lipsum}
\usepackage{multirow}
\usepackage{nicefrac}
\usepackage[normalem]{ulem}
\usepackage[dvipsnames]{xcolor}
\usepackage{hyperref}
\hypersetup{
    colorlinks=true,
    urlcolor=blue,
    linkcolor = blue,
    urlcolor  = blue,
    citecolor = blue,
    anchorcolor = blue,
    pdftitle={JBW-CC-electroproduction},
    pdfauthor={JBW},
    pdfsubject={JBW-CC-electroproduction}}

\allowdisplaybreaks
\usepackage{cleveref}
\crefformat{figure}{Fig.~#2#1#3}
\crefformat{table}{Tab.~#2#1#3}
\crefformat{equation}{Eq.~(#2#1#3)}
\crefformat{section}{Sect.~#2#1#3}
\begin{document}

%%%%%%%%%%%%%%%%%%%%%%%%%%%%%%%%%%%%%%%%%%%%%%%

\title{
Testing chiral unitary models for the \texorpdfstring{$\Lambda(1405)$}{} in  \texorpdfstring{$K^{+}\pi\Sigma$}{} photoproduction}

\author{P.~C.~Bruns}
\email{bruns@ujf.cas.cz}
\affiliation{Nuclear Physics Institute of the Czech Academy of Sciences, 250 68 \v{R}e\v{z}, Czechia}

\author{A.~Ciepl\'y}
\email{cieply@ujf.cas.cz}
\affiliation{Nuclear Physics Institute of the Czech Academy of Sciences, 250 68 \v{R}e\v{z}, Czechia}

\author{M.~Mai}
\email{mai@hiskp.uni-bonn.de}
\affiliation{Helmholtz-Institut f\"ur Strahlen- und Kernphysik (Theorie) and Bethe Center for Theoretical Physics, Universit\"at Bonn, D--53115 Bonn, Germany}
\affiliation{The George Washington University, Washington, DC 20052, USA}

%%%%%%%%%%%%%%%%%%%%%%%%%%%%%%%%%%%%%%%%%%%%%%%%%%%%%%%
%%%%%%%%%%%%%%%%%%%%%%%%%%%%%%%%%%%%%%%%%%%%%%%%%%%%%%%
\begin{abstract}
We adopt a novel formalism for the low-energy analysis of the $\gamma p \to K^{+}\pi\Sigma$ photoproduction reaction to calculate the $\pi\Sigma$ invariant mass distributions in the $\Lambda(1405)$ resonance region. The approach adheres to constraints arising from unitarity, gauge invariance and chiral perturbation theory, and is used without adjusting any model parameters. It is found that the meson-baryon rescattering in the final state has a major impact on the magnitude and structure of the generated spectra that are compared with the experimental data from the CLAS collaboration. We demonstrate a large sensitivity of the theoretical predictions to the choice of the coupled-channel $\pi\Sigma - \bar{K}N$ model amplitudes which should enable one to constrain the parameter space of these models.
\end{abstract}
%%%%%%%%%%%%%%%%%%%%%%%%%%%%%%%%%%%%%%%%%%%%%%%%%%%%%%%
%%%%%%%%%%%%%%%%%%%%%%%%%%%%%%%%%%%%%%%%%%%%%%%%%%%%%%%
%%%%%%%%%%%%%%%%%%%%%%%%%%%%%%%%%%%%%%%%%%%%%%%%%%%%%%%
\maketitle   %%%%%%%%%%%%%%%%%%%%%%%%%%%%%%%%%%%%%%%%%%
%%%%%%%%%%%%%%%%%%%%%%%%%%%%%%%%%%%%%%%%%%%%%%%%%%%%%%%

%%%%%%%%%%%%%%%%%%%%%
%%%%%%%%%%%%%%%%%%%%%
\begin{figure*}[t]
\centering
\subfigure[\,Weinberg-Tomozawa (WT) graphs.]
{
\begin{minipage}{0.25\linewidth}
\includegraphics[width=\textwidth]{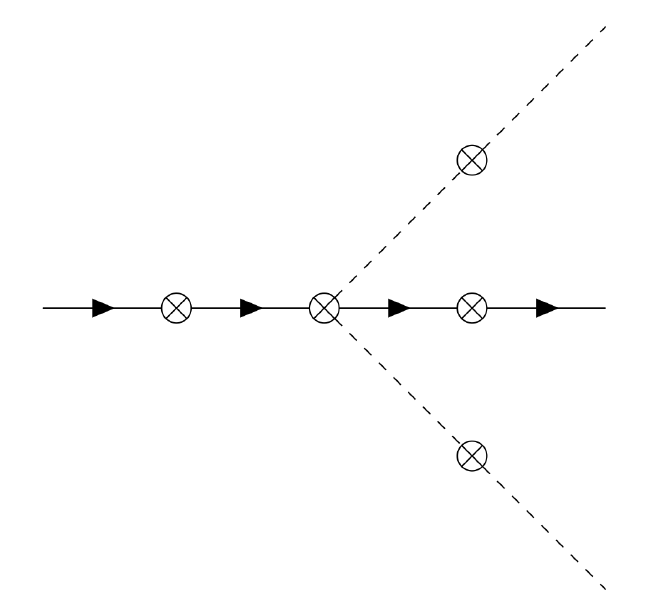}
\end{minipage}
}
\subfigure[\,Born term (BT) graphs of two types, B1 and B2.]
{
\begin{minipage}{0.35\linewidth}
\includegraphics[width=\textwidth]{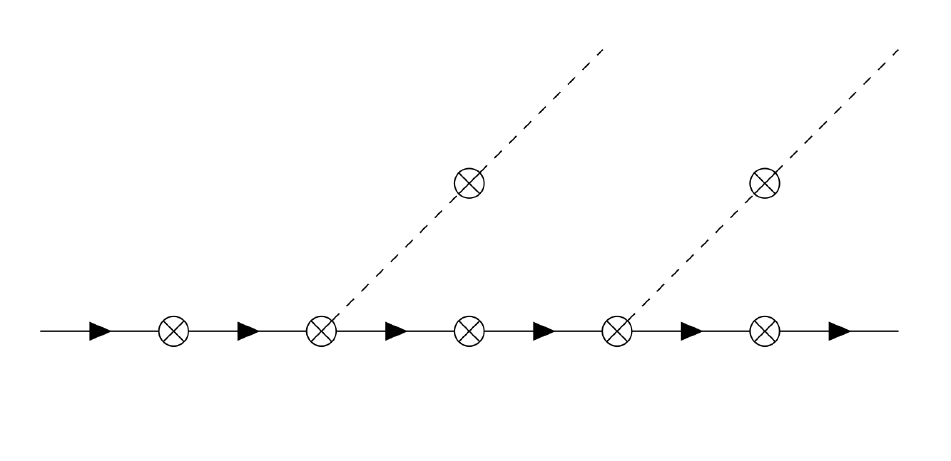}\\
\vspace{1.2cm}
\end{minipage}
}
\subfigure[\,The anomalous (AN) graph.]
{
\begin{minipage}{0.23\linewidth}
\includegraphics[width=\textwidth]{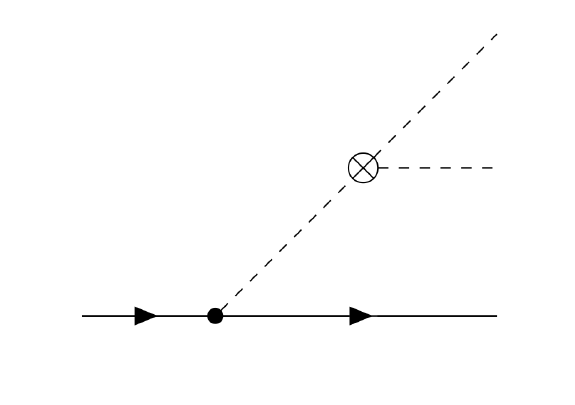}\\
\vspace{1.4cm}
\end{minipage}
}
\caption{Three classes of tree graphs for the two-meson photoproduction amplitude. Directed lines: baryons, 
dashed lines: pseudoscalar mesons, crossed circles: possible photon insertions. Working at order $\mathcal{O}(e)$, 
the first figure (a) represents five graphs, each with the photon attached to a different crossed vertex. 
The same notation system applies to figures (b) and (c) giving altogether rise to $5+(2\times7)+1=20$ tree graphs.}
\label{fig:tree_graphs}
\end{figure*}
%%%%%%%%%%%%%%%%%%%%%
%%%%%%%%%%%%%%%%%%%%%

%%%%%%%%%%%%%%%%%%%%%%%%%%%%%%%%%%%%%%%%%%%%%%%%%%%%%%%%%%%%%%%%%%%%%%%%%%%%%%%%%%%%%%%%%%%%%%%
\section{Introduction}
\label{sec:introduction}
%%%%%%%%%%%%%%%%%%%%%%%%%%%%%%%%%%%%%%%%%%%%%%%%%%%%%%%%%%%%%%%%%%%%%%%%%%%%%%%%%%%%%%%%%%%%%%%

Unraveling the general patterns and the nature of individual (excited) states in the hadron spectrum is essential to test our understanding of the strong interaction. In that, only few other states have such 
a long-lasting history as the enigmatic $\Lambda(1405)$ resonance, becoming a benchmark for our understanding of the $\rm SU(3)$ hadron dynamics. Starting with its theoretical prediction \cite{Dalitz:1959dn} and later 
experimental verification \cite{Alston:1961}, until the most recent debates on the existence of the second broad pole, 
it emerged as a fruitful research area sparking many theoretical and experimental developments. 
For a deeper recap of the character and history of $\Lambda(1405)$, established to be an $I(J^P)=0\,(1/2^-)$ 
resonance in the strangeness $S=-1$ sector, we refer the reader to the recent dedicated reviews~\cite{Hyodo:2011ur,Mai:2020ltx}. More general reviews of our present understanding of excited hadrons 
can be found in Refs.~\cite{Hyodo:2020czb,Mai:2022eur}.

Broader impact of the research related to the antikaon-nucleon systems includes:
(1) Application to the $\Lambda_b\to J/\psi\, \Lambda(1405)$ decay~\cite{Roca:2015tea} and similar processes 
with the final-state interaction dominated by the non-perturbative meson-baryon dynamics~\cite{Oset:2016lyh}; 
(2) Investigation of and search for $\bar K$-nuclei ($\bar KNN$, $\bar KKN$, $\bar KNNN$, etc.), 
being part of large experimental programs at DA$\rm \Phi$NE~\cite{Agnello:1998vc, Agnello:2005qj}, 
Saclay~\cite{Yamazaki:2008hm, Maggiora:2009gh, Yamazaki:2010mu} or at J-PARC \cite{J-PARCE15:2016esq,J-PARCE15:2020gbh}
to name just a few; (3) The exploration of the in-medium properties of anti-kaons and strange nuclear 
matter~\cite{Cieply:2011fy, Gal:2014uua, Hrtankova:2018fjo}. One natural application is also represented 
by studies of the equation of state of neutron stars in relation to the strangeness, see the comprehensive 
review~\cite{Gal:2016boi} on many aspects related to strangeness in nuclear physics.

The state-of-the-art theoretical approaches to the universal parameters of the sub-threshold  $\Lambda(1405)$ 
resonance include various forms of coupled-channel models implementing $S$-matrix unitarity, while also adding constraints from chiral symmetry of QCD to various degrees. Parameters of such models are typically fixed by fitting $K^{-}p$ reactions and threshold data, and the chirally motivated coupled-channel dynamics generates two poles in the complex energy plane~\cite{Oller:2000fj, Jido:2003cb, Mai:2012dt,Cieply:2016jby}. 
The theoretical predictions \cite{Ikeda:2012au,Guo:2012vv,Mai:2014xna,Feijoo:2018den,Bruns:2021krp} tend to agree on the position 
of the narrow pole located around ${1425~{\rm MeV}}$, coupling strongly to the $\bar KN$ channel and mostly interpreted 
as a meson-baryon molecular state. The position of the second pole is not determined so well, is broader and at significantly lower energies. This so-called \emph{double-pole} structure of $\Lambda(1405)$ is what makes this state particularly curious, while we know now that such a structure is common to many other states 
in the hadron spectrum~\cite{Meissner:2020khl,Mai:2022eur}.

One particular complication in accessing the parameters of both $\Lambda(1405)$ states directly 
lies in the fact that there is no $\Sigma$ target to perform $\pi\Sigma$ scattering experiments. 
This hurdle can be overcome by analyzing processes with $\pi\Sigma$ appearing in the final state.
This is the case, e.g., for the \emph{two-meson photoproduction} reaction $\gamma p \to K^{+}\pi\Sigma$ 
in which the $K^+$ meson takes away momentum, enabling a scan in the invariant mass of the $\pi\Sigma$ system 
down to its production threshold. Such determined line-shape of the $\pi\Sigma$ system covers the relevant 
energy and, in principle, allows one to extract the parameters of the $S=-1$ meson-baryon system more directly. 
The corresponding experiment was performed by the CLAS collaboration several years ago, taking very precise, high-resolution 
data~\cite{Moriya:2013eb, Schumacher:2013vma, CLAS:2013rxx}. These data have already made a significant impact on the field, see 
Refs.~\cite{Roca:2013cca,Mai:2014xna,Roca:2013av,Nakamura:2013boa,Wang:2016dtb,Nam:2017yeg,Anisovich:2020lec,Anisovich:2021wct},
where various photoproduction models have been tested and the generated $\pi\Sigma$ mass distributions compared with the data. 
Our work aims to contribute to these efforts by implementing a recently proposed formalism~\cite{Bruns:2020lyb} 
for the $\gamma p \to K^{+}\pi\Sigma$ transition and discussing the sensitivity of the calculated cross 
sections to ambiguities inherent in extending the $\pi\Sigma - \bar{K}N$ coupled-channel models to energies below the $\bar{K}N$ 
threshold. In the future, this should allow us to reduce the model dependence and shrink at the same time the parameter space.

At this point, it might be appropriate to mention other contemporary and ongoing experimental studies aiming at shedding 
more light on this enigmatic part of the hadron spectrum. First of all, the $\pi\Sigma$ mass spectra were also measured 
in other reactions, some of them compatible with the CLAS data \cite{Zychor:2007gf,BGOOD:2021sog}, some other not \cite{HADES:2012csk}. 
Completely new experimental data on the isoscalar $\pi^{0}\Sigma^{0}$ photoproduction should also be reported soon 
by the GlueX collaboration \cite{2022GlueX}. Other efforts aim at determining the isovector part of the $\bar{K}N$ 
interaction via precise measurements of the kaonic hydrogen and kaonic deuterium characteristics 
\cite{SIDDHARTA:2013ftj,Curceanu:2013bxa,Marton:2016pwv} or by studying the related $K^{0}_L p \to K^+\Xi^0$ 
reaction~\cite{KLF:2020gai}. Finally, exciting new prospects in exploring strangeness in the hadron spectrum 
may open with the PANDA experiment~\cite{PANDA:2009yku,PANDA:2021ozp}.

The article is organized as follows: In \cref{sec:formalism} we introduce the formalism adopted to study the $K^{+}\pi\Sigma$ 
photoproduction on proton targets. In \cref{sec:expdata} we discuss the predictions made using this formalism, and compare 
the resulting $\pi\Sigma$ mass distributions with the CLAS data. Finally, in \cref{sec:conclusion} we conclude summarizing 
our findings and provide an outlook for future directions.

%%%%%%%%%%%%%%%%%%%%%%%%%%%%%%%%%%%%%%%%%%%%%%%%%%%%%%%%%%%%%%%%%%%%%%%%%%%%%%%%%%%%%%%%%%%%%%%
\section{Formalism}
\label{sec:formalism}
%%%%%%%%%%%%%%%%%%%%%%%%%%%%%%%%%%%%%%%%%%%%%%%%%%%%%%%%%%%%%%%%%%%%%%%%%%%%%%%%%%%%%%%%%%%%%%%

We consider the two-meson photoproduction reaction $\gamma(k)p(p_{N})\,\rightarrow\,K(q_{K})\pi(q_{\pi})\Sigma(p_{\Sigma})$\,, 
where the symbol in brackets denotes the four-momentum of the indicated particle. For a process with five external particles, 
we can form five independent Mandelstam variables, which we choose here as
\begin{align} 
  s               &= (p_{N}+k)^2 = (q_{K}+q_{\pi}+p_{\Sigma})^2\,,                  \nonumber \\
  M_{\pi\Sigma}^2 &= (q_{\pi}+p_{\Sigma})^2\,, \quad t_{\Sigma} = (p_{\Sigma}-p_{N})^2\,, \nonumber \\
  u_{\Sigma}      &= (p_{\Sigma}-k)^2\,, \quad \;t_{K} = (q_{K}-k)^2\,.
  \label{eq:mandelstams}
\end{align}

The two-meson photoproduction amplitude can be decomposed as 
\begin{align}
\label{eq:MmuDecomp}
\mathcal{M}^{\mu} =& \,\gamma^{\mu}\mathcal{M}_{1} + p_{N}^{\mu}\mathcal{M}_{2} + p_{\Sigma}^{\mu}\mathcal{M}_{3} + q_{K}^{\mu}\mathcal{M}_{4}   \\ 
&+ \slashed{k}\left(\gamma^{\mu}\mathcal{M}_{5} + p_{N}^{\mu}\mathcal{M}_{6} + p_{\Sigma}^{\mu}\mathcal{M}_{7} + q_{K}^{\mu}\mathcal{M}_{8} \right) \nonumber \\ 
&+ \slashed{q}_{K}\left(\gamma^{\mu}\mathcal{M}_{9} + p_{N}^{\mu}\mathcal{M}_{10} + p_{\Sigma}^{\mu}\mathcal{M}_{11} + q_{K}^{\mu}\mathcal{M}_{12} \right) \nonumber \\ 
&+ \slashed{q}_{K}\slashed{k}\left(\gamma^{\mu}\mathcal{M}_{13} + p_{N}^{\mu}\mathcal{M}_{14} + p_{\Sigma}^{\mu}\mathcal{M}_{15} + q_{K}^{\mu}\mathcal{M}_{16} \right)\,,\nonumber
\end{align}
where we have suppressed structures $\sim k^{\mu}$ that do not contribute to the photoproduction cross section. 
The structures $\mathcal{M}$ can be obtained diagrammatically as it is shown explicitly in App.~C of Ref.~\cite{Bruns:2020lyb}. 
At the leading order, there are 20 tree level diagrams contributing to these structures, which can be arranged in three classes: (WT) graphs with one Weinberg-Tomozawa contact term with a photon attached to one of five possible places; (B1/B2) Born graphs with the pion/kaon emitted before the kaon/pion, respectively, including seven possible photon coupling possibilities; (AN) graphs with a three-meson-photon vertex from the anomalous Wess-Zumino-Witten Lagrangian \cite{Wess:1971yu,Witten:1983tw}. Thus, $5+(2\times7)+1=20$ tree diagrams comprise our photoproduction kernel.
We also remark that the latter class of graphs yields contributions of sub-leading chiral order in the low-energy power counting, but we include it here as a test of higher-order contributions, without introducing new adjustable parameters. 

Our notation for kinematics is such that Lorentz-non-invariant quantities (like angles and three-momenta) 
are marked with a $(\ast)$ if they are evaluated in the $\pi\Sigma$ c.m. frame (where $\vec{p}_{\Sigma}^{\,\ast}+\vec{q}_{\pi}^{\,\ast}=\vec{0}$), and otherwise refer to the overall c.m. frame where $\vec{p}_{N}+\vec{k}=\vec{0}$. For example, we have
\begin{align}
|\vec{q}_{K}| &= \frac{\sqrt{\lambda(s,M_{\pi\Sigma}^2,M_{K}^2)}}{2\sqrt{s}}\,,\label{eq:qK}\\
|\vec{p}_{\Sigma}^{\,\ast}| &= \frac{\sqrt{\lambda(M_{\pi\Sigma}^2,m_{\Sigma}^2,M_{\pi}^2)}}{2M_{\pi\Sigma}}\,,\label{eq:pSstar}
\end{align}
employing $\lambda(x,y,z):=x^2+y^2+z^2-2xy-2xz-2yz$\,.\\

In the coupled-channel formalism of \emph{Unitarized ChPT}, one constructs meson-baryon partial-wave 
scattering amplitudes $f_{\ell\pm}^{c',c}(M_{\pi\Sigma})$, which aim to describe the scattering 
from channel $c$ to channel $c'$ (here, $c, c'=\pi\Sigma,\,\bar{K}N,\,\eta\Lambda,\ldots$) 
for total angular momentum $\ell\pm\frac{1}{2}$ and orbital angular momentum \mbox{$\ell=0,1,2,\ldots\,$}. 
These amplitudes are designed in such a way to be consistent with ChPT up to a fixed order 
of the low-energy expansion (in practice, usually on tree level, 
i.e., $\mathcal{O}(p)$ or $\mathcal{O}(p^2)$). At the same time, they fulfill the requirement of coupled-channel unitarity, 
%%%%%%%%%%%%%%%%%%%%%
\begin{equation}
\label{eq:unif}
\mathrm{Im}(f_{\ell\pm}) = (f_{\ell\pm})^{\dagger}(|\vec{p}^{\,\ast}|)(f_{\ell\pm}) \;,
\end{equation}
%%%%%%%%%%%%%%%%%%%%%
above the lowest reaction threshold. Here, $f_{\ell\pm}$ denotes a complex-valued matrix in the space 
of the considered meson-baryon channels $c,c'$, with entries $f_{\ell\pm}^{c',c}(M_{\pi\Sigma})$, 
while $(|\vec{p}^{\,\ast}|)$ is a diagonal matrix in this space, the entries of which are given 
by the moduli of the three-momenta (for each channel $c$) in the meson-baryon c.m.~frame 
(see \cref{eq:pSstar}) above the threshold of channel $c$, and zero below it.

%%%%%%%%
%%%%%%%%
\begin{figure}[t]
\centering
\includegraphics[width=0.5\textwidth]{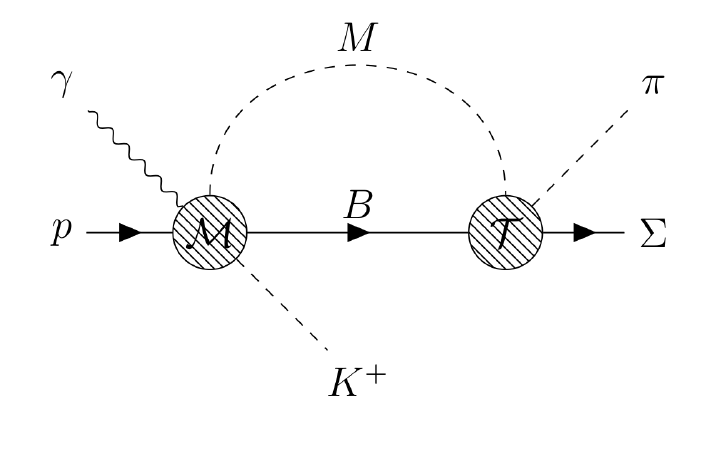}
\caption{Representation of the final-state interaction (FSI) of the $S=-1$ meson-baryon pair $MB$. $\mathcal{M}$ is the amplitude for $\gamma p\rightarrow K^{+}MB$ without FSI, and $\mathcal{T}$ is the $S=-1$ meson-baryon scattering amplitude, which can be decomposed into partial waves $f_{\ell\pm}$ \cite{Chew:1957zz}.}
\label{fig:MBFSI}
\end{figure}
%%%%%%%%
%%%%%%%%

It is our aim to implement the scattering amplitudes $f_{0+}$ in the photoproduction formalism 
in such a way, that it describes the final-state s-wave interaction of the $S=-1$ meson-baryon 
pair $(MB)$ produced in the $\gamma p\rightarrow K^{+}MB$ reaction as illustrated in  Fig.~\ref{fig:MBFSI}. 
Clearly, we have to find the combinations of (partial-wave projections of) the structure functions 
$\mathcal{M}_{i=1,\ldots,16}$ which project on the $\ell=0$ state of this meson-baryon pair. 
In fact, one can construct projected photoproduction amplitudes $\mathcal{A}^{i=1,\ldots,4}_{0+}(s,M_{\pi\Sigma}^2,t_{K})$ 
from the $\mathcal{M}_{i}$ which have simple unitarity relations with the pertinent s-wave scattering amplitudes,
%%%%%%%%%%%%%%%%%%%%%
\begin{equation}
\label{eq:uniA}
\mathrm{Im}(\mathcal{A}^{i}_{0+}) = (f_{0+})^{\dagger}(|\vec{p}^{\,\ast}|)(\mathcal{A}^{i}_{0+})\,,\quad i=1,\ldots,4\,.
\end{equation}
%%%%%%%%%%%%%%%%%%%%%
Note that there are \emph{four} amplitudes for the production of a state with $\pi\Sigma$ being in an s-wave, corresponding to the four invariant amplitudes for the photoproduction process $\gamma p\rightarrow K^{+}\Lambda^{\ast}$, see Ref.~\cite{Bruns:2020lyb} for more details. The projections $\mathcal{A}^{i}_{0+}$ can be found, e.g., by applying the Cutkosky rules \cite{Cutkosky:1960sp} 
to the $MB$ loop in Fig.~\ref{fig:MBFSI}, and studying the consequences of unitarity for the various invariant structures in $\mathcal{M}^{\mu}$, or equivalently by the methods used in Ref.~\cite{Chew:1957tf,Pearlstein:1957zz,Berends:1967vi} to extract the multipole amplitudes for single-meson photoproduction. Explicit expressions are provided in Appendix \ref{app:cc_formalism}.  This analysis is somewhat simplified employing an approximation which is motivated by the observations made in Sec.~4 of Ref.~\cite{Bruns:2020lyb}. In that, confining ourselves to a low-energy analysis in the $\Lambda(1405)$ energy region, it is reasonable to neglect higher partial waves $\sim\mathcal{Y}_{\ell>0, m}(\theta_{\Sigma}^{\ast},\phi_{\Sigma}^{\ast})$ in the decomposition of the $\mathcal{M}_{i}$ into spherical harmonics, and express the projections solely through
%%%%%%%%%%%%%%%%%%%%%
\begin{equation}\label{eq:MbarApprox}
\overline{\mathcal{M}}_{i}(s,M_{\pi\Sigma}^2,t_{K}):=\int\frac{d\Omega_{\Sigma}^{\ast}}{4\pi}\mathcal{M}_{i}(s,M_{\pi\Sigma}^2,t_{K},t_{\Sigma},u_{\Sigma})\,. 
\end{equation}
%%%%%%%%%%%%%%%%%%%%%
The s-wave resonance pole terms in the photoproduction amplitude are not affected by this approximation. We also note that this approximation can be used for any c.m. energy $\sqrt{s}$, as long as $M_{\pi\Sigma}$ stays sufficiently close to the $MB$ threshold region. Therefore, kinematics involving high-energy kaons can, in principle, be treated within the framework of this section. Of course, the ChPT tree graphs might not be sufficient for this purpose, and the elementary photoproduction amplitude would have to be amended.

Neglecting contributions due to the $\pi\Sigma$ states with $\ell>0$, the double-differential cross section 
for the $\gamma p\rightarrow K^{+}\pi\Sigma$ reaction can be expressed through the $\mathcal{A}^{i}_{0+}$ 
amplitudes as follows,
%%%%%%%%%%%%%%%%%%%%%
\begin{equation}
\label{eq:d2csA}
 \frac{d^2\sigma}{d\Omega_{K}dM_{\pi\Sigma}} = \frac{|\vec{q}_{K}||\vec{p}_{\Sigma}^{\,\ast}|}{(4\pi)^4s|\vec{k}|}|\mathcal{A}|^2\,,
\end{equation}
%%%%%%%%%%%%%%%%%%%%%
where we introduced
%%%%%%%%%%%%%%%%%%%%% 
\begin{align*}
&4|\mathcal{A}|^2 = (1\!-\!z_{K})\left|\mathcal{A}_{0+}^{1} + \mathcal{A}_{0+}^{2}\right|^2 
                 + (1\!+\!z_{K})\left|\mathcal{A}_{0+}^{1} - \mathcal{A}_{0+}^{2}\right|^2 \\
&\hspace*{5.5ex} + (1\!-\!z_{K})\,\biggl|\mathcal{A}_{0+}^{1} + \mathcal{A}_{0+}^{2}  \\ 
&+\frac{2|\vec{q}_{K}|(1+z_{K})}{M_{K}^2-t_{K}} \left((\sqrt{s}+m_{N})\mathcal{A}_{0+}^{3} + (\sqrt{s}-m_{N})\mathcal{A}_{0+}^{4}\right)\biggr|^2 \\  
&\hspace*{5.5ex} + (1\!+\!z_{K})\,\biggl|\mathcal{A}_{0+}^{1} - \mathcal{A}_{0+}^{2}  \\
&-\frac{2|\vec{q}_{K}|(1\!-\!z_{K})}{M_{K}^2-t_{K}} \left((\sqrt{s}+m_{N})\mathcal{A}_{0+}^{3} - (\sqrt{s}-m_{N})\mathcal{A}_{0+}^{4}\right)\biggr|^2 \,, 
\end{align*}
%%%%%%%%%%%%%%%%%%%%%
with $z_{K}\equiv\cos\theta_{K}$, $\theta_{K}$ being the angle between $\vec{q}_{K}$ 
and $\vec{k}$ in the overall c.m.~frame. 

Returning to the partial-wave unitarity statement in \cref{eq:uniA}, the proposed coupled-channel 
formalism is now easily explained. We have to construct an elementary photoproduction amplitude, 
e.g., from the tree graphs discussed above, computing the according projected amplitudes 
$\mathcal{A}^{i(\mathrm{tree})}_{0+}$, as detailed. Since we treat the outgoing $K^{+}$ effectively 
as a spectator particle in our unitarization procedure, as was also done in previous studies \cite{Nacher:1998mi,Lutz:2004sg,Nakamura:2013boa,Roca:2013cca,Roca:2013av,Mai:2014xna}, the final-state interaction 
in our model is restricted to the $S=-1$ meson-baryon subspace, which is the relevant sector for the formation 
of the $\Lambda(1405)$.
Therefore, unitarized amplitudes for $\gamma p\rightarrow K^{+}MB$ will be taken as the coupled-channel vector
%%%%%%%%%%%%%%%%%%%%%
\begin{equation}
\label{eq:modelA}
(\mathcal{A}^{i}_{0+}) = (\mathcal{A}^{i(\mathrm{tree})}_{0+}) + (f_{0+})\left(8\pi M_{\pi\Sigma}G(M_{\pi\Sigma})\right)(\mathcal{A}^{i(\mathrm{tree})}_{0+})\,.
\end{equation}
%%%%%%%%%%%%%%%%%%%%%
Here, $G(M_{\pi\Sigma})$ is a diagonal channel-space matrix, with entries given by suitably regularized loop integrals
%%%%%%%%%%%%%%%%%%%%%
\begin{align}
\mathrm{i}\, & G^{c=MB}(M_{\pi\Sigma}) = \\ &\int_{\mathrm{reg.}}\frac{d^{4}l}{(2\pi)^4}\frac{1}{((p_{\Sigma}+q_{\pi}-l)^2-m_{B}^2+i\epsilon)(l^2-M_{M}^2+i\epsilon)}
 \,. \nonumber
\end{align}
%%%%%%%%%%%%%%%%%%%%%
The $MB=\pi\Sigma$ entries of the vector $\mathcal{A}^{i}_{0+}$ can then be inserted in formula \cref{eq:d2csA} to obtain the required s-wave cross sections. Using the fact that
%%%%%%%%%%%%%%%%%%%%%
\begin{equation}\label{eq:ImG}
  8\pi M_{\pi\Sigma}\,\mathrm{Im}\,G^{MB}(M_{\pi\Sigma})=|\vec{p}_{B}^{\,\ast}|\Theta(M_{\pi\Sigma}-m_{B}-M_{M})\,,
\end{equation}
%%%%%%%%%%%%%%%%%%%%%
where $\Theta(\cdot)$ denotes the Heaviside step function, together with Eq.~(\ref{eq:unif}), it is straightforward to show 
that the ansatz (\ref{eq:modelA}) solves the unitarity requirement (\ref{eq:uniA}).

In the actual calculations presented in the next section we will use $f_{0+}$ amplitudes generated by two different 
approaches to $MB$ coupled-channel interactions. Both of them derive the interaction kernel from the $MB$ chiral Lagrangian 
taken up to the NLO order, but differ (among other details) in the methods adopted to regularize the loop function $G$. 
In order to stay consistent with the construction of the amplitudes, we will use in each case the appropriate regulation 
procedure that is employed in the respective model. Thus, in the case of the Bonn models presented in Refs.~\cite{Mai:2014xna} 
and \cite{Sadasivan:2018jig}, we take the dimensionally regularized integral $G=-I_{MB}$ 
(see App.~B of the first reference), while in the case of the Prague model~\cite{Bruns:2021krp}, we take 
the loop integral of Eq.~(8) in \cite{Bruns:2019fwi}, divided by $8\pi M_{\pi\Sigma}\,g^2_{jb}$, to conform 
with Eq.~(\ref{eq:ImG}). Requiring that both versions of the loop function $G$ agree at the threshold energy $M_{MB}=m_B+M_M$
yields a matching relation between the regulator scale $\alpha$ used in  Refs.~\cite{Bruns:2021krp,Bruns:2019fwi} 
and the mass scale $\mu$ adopted in Refs.~\cite{Mai:2014xna, Sadasivan:2018jig}:
%%%%%%%%%%%%%%%%%%%%%
\begin{equation}\label{eq:regmatch}
\pi \alpha\overset{!}{=} m_B+M_M-m_B\log\left(\frac{m_B^2}{\mu^2}\right) - M_M\log\left(\frac{M_M^2}{\mu^2}\right)\,. 
\end{equation}
%%%%%%%%%%%%%%%%%%%%%
For a \emph{natural} scale of $\mu \approx 1~{\rm GeV}$ used in dimensional regularization this somewhat ad-hoc matching 
procedure provides $\alpha_{\pi\Sigma} \approx 460$ MeV and $\alpha_{\bar{K}N} \approx 715$ MeV, both values 
obtained via Eq.~(\ref{eq:regmatch}) at the respective channel thresholds.

After ensuring two-body unitarity in the final state we turn to another aspect, the gauge invariance of the photoproduction 
amplitude. Could there be a conflict between Eqs.~(\ref{eq:d2csA})-(\ref{eq:modelA}) and gauge invariance? 
The answer is not trivial. Even a gauge-invariant tree-level amplitude $\mathcal{M}^{\mu}_{(\mathrm{tree})}$ does not 
necessarily generate a gauge-invariant unitarized amplitude when it is just plugged into a loop integral to couple it 
to the final-state interaction, since it will usually depend on the loop momentum that is integrated over. 
We refer here to the discussions in \cite{vanAntwerpen:1994vh,Borasoy:2005zg,Borasoy:2007ku,Mai:2012wy}, 
see also  Refs.~\cite{Haberzettl:2018hcl,Haberzettl:2021wcz}. In our present framework, the issue is resolved as follows. 
Given the functions $\mathcal{A}^{i}_{0+}(s,M_{\pi\Sigma}^2,t_{K})$, we can find a set of invariant amplitudes $\mathcal{M}_{i}^{\mathcal{C}}$ (provided in App.~\ref{app:cc_formalism}) which form a gauge-invariant amplitude by construction, and yield the same projections $\mathcal{A}^{i}_{0+}$. The s-wave cross-section 
calculated directly from this gauge-invariant amplitude exactly equals the one in Eq.~(\ref{eq:d2csA}). 
The difference between the invariant amplitudes, thus, resides in the higher meson-baryon partial waves.
This strategy works for every gauge-invariant set of $\mathcal{A}^{i(\mathrm{tree})}_{0+}$ 
in (\ref{eq:modelA}), which allows for some flexibility in the construction of models for these functions. For example, we could add higher-order contact terms to the ChPT tree graphs. All these models will yield cross-sections that are in accord with s-wave coupled-channel unitarity, gauge invariance and the chiral low-energy theorems~\cite{Bruns:2020lyb} (as long as the model used for the $f_{0+}$ does not spoil 
the proper low-energy behaviour) at the same time.

%%%%%%%%
%%%%%%%%
\begin{figure}[htb]
    \includegraphics[width=0.42\textwidth]{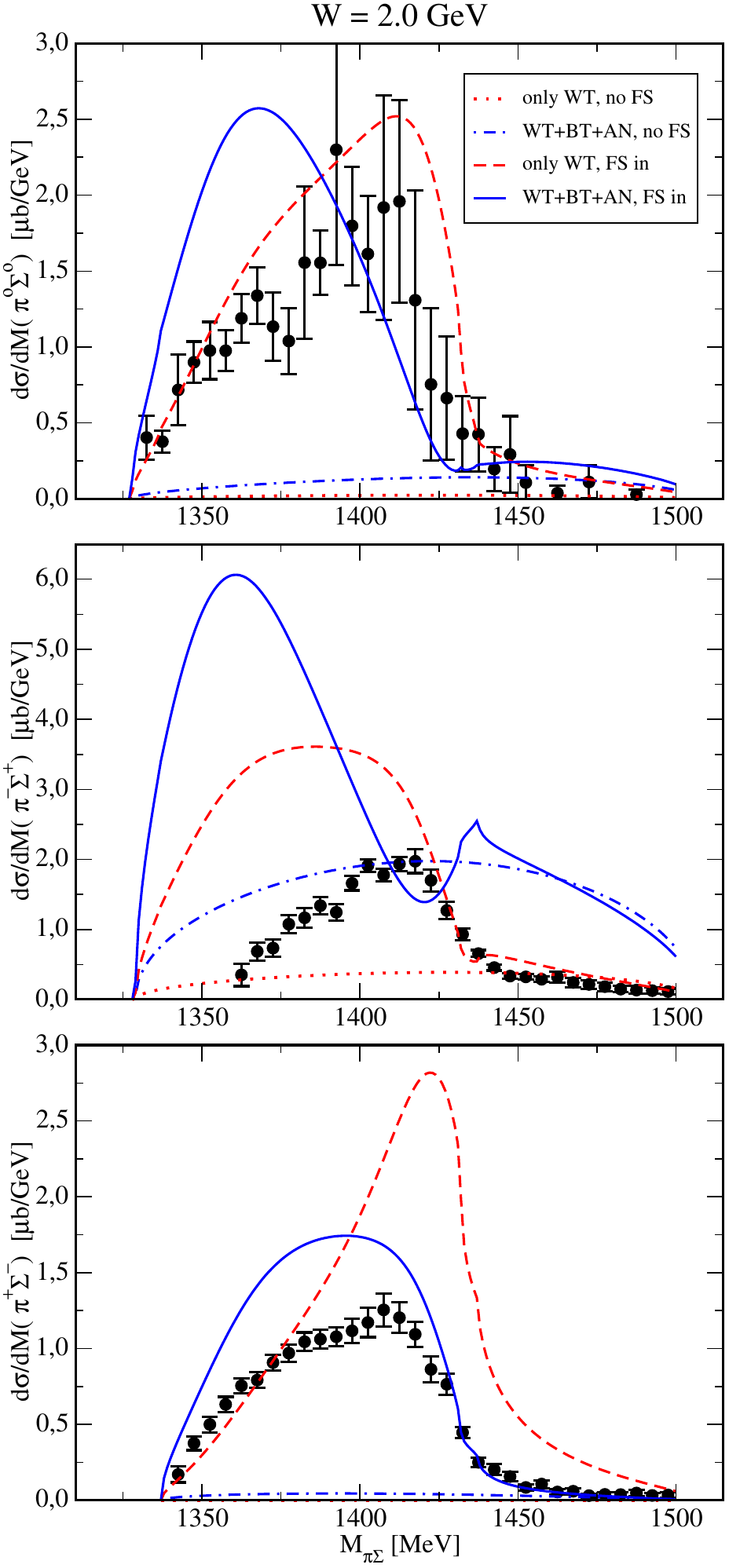}
    \vspace*{-3mm}
    \caption{The calculated $\pi\Sigma$ mass distributions are compared with experimental data taken from \cite{Moriya:2013eb}. 
    The lines demonstrate the contributions from WT and other (BT+AN) graphs, and the impact of the $MB$ rescattering in the final state (FS).}
    \label{fig:FSimpact}
\end{figure}
%%%%%%%%
%%%%%%%%

%%%%%%%%%%%%%%%%%%%%%%%%%%%%%%%%%%%%%%%%%%%%%%%%%%%%%%%%%%%%%%%%%%%%%%%%%%%%%%%%%%%%%%%%%%%%%%%
\section{Comparison with data}
\label{sec:expdata}
%%%%%%%%%%%%%%%%%%%%%%%%%%%%%%%%%%%%%%%%%%%%%%%%%%%%%%%%%%%%%%%%%%%%%%%%%%%%%%%%%%%%%%%%%%%%%%%

We use the formalism described in the previous section to calculate the $\pi\Sigma$ mass distributions 
observed in the CLAS experiment \cite{Moriya:2013eb} for the photoproduction reactions on a proton target. 

First, we look at the relevance of various contributions to the s-wave amplitude of the photoproduction 
process. In Fig.~\ref{fig:FSimpact} we present the results obtained for the mass spectra at $W=\sqrt{s}=2.0~{\rm GeV}$ 
when only the tree level graphs contribute to the amplitude, and demonstrate the additional impact of including 
the $MB$ rescattering in the final state. For the latter, we have adopted only the recent version of 
the $\bar{K}N$ Prague model \cite{Bruns:2021krp} to generate the rescattering amplitudes $f_{0+}$ 
required in Eq.~(\ref{eq:modelA}), but will present results including more models below.
We observe that when only the WT tree graph is considered (dotted red lines in the figure), 
without the rescattering second term of Eq.~(\ref{eq:modelA}), the $\pi\Sigma$ cross sections are either 
exactly zero (for $\pi^{+}\Sigma^{-}$)\footnote{This result is caused by a structure of the WT graph in this case.}, 
quite negligible (for $\pi^{0}\Sigma^{0}$) or remain very small (for $\pi^{-}\Sigma^{+}$). 
The addition of Born and anomalous terms leads to larger cross sections (shown by dot-dashed blue 
lines in the figure) and two of the predicted $\pi\Sigma$ spectra compare reasonably well with those observed in the CLAS experiment~\cite{Moriya:2013eb}. The exception here is the $\pi^{-}\Sigma^{+}$ distribution
that does not match the data neither in magnitude, nor in its shape. 
It should be noted that all three $\pi\Sigma$ mass distributions generated with photoproduction amplitudes constructed from tree level graphs are very flat. The peak structure appears only when the $MB$ rescattering is taken into account in the final state, as the dashed red (for only WT graph)
and continuous blue lines (for all graphs) show. In general, the inclusion of the Born terms moves the peak structure to lower energies. We have also checked that the contribution of the anomalous graphs is relatively 
small. For this reason, we do not show their impact on the spectra separately to avoid overcrowding 
the figures with too many lines. We also conclude that at least at the c.m.~energy $W=2.0~{\rm GeV}$ 
the description of the $\pi^{0}\Sigma^{0}$ and $\pi^{+}\Sigma^{-}$ cross sections is quite reasonable. 
Although we cannot say the same about the $\pi^{-}\Sigma^{+}$ mass spectrum, one should bear in mind 
that our theoretical predictions are provided without any adjustment of the $MB$ rescattering amplitudes 
that have a major impact on the calculated spectra but are generated by a model fitted to a completely different 
sector of experimental data and for much higher energies from the $\bar{K}N$ threshold up.

\begin{figure*}[t]
    \includegraphics[width=0.8\textwidth]{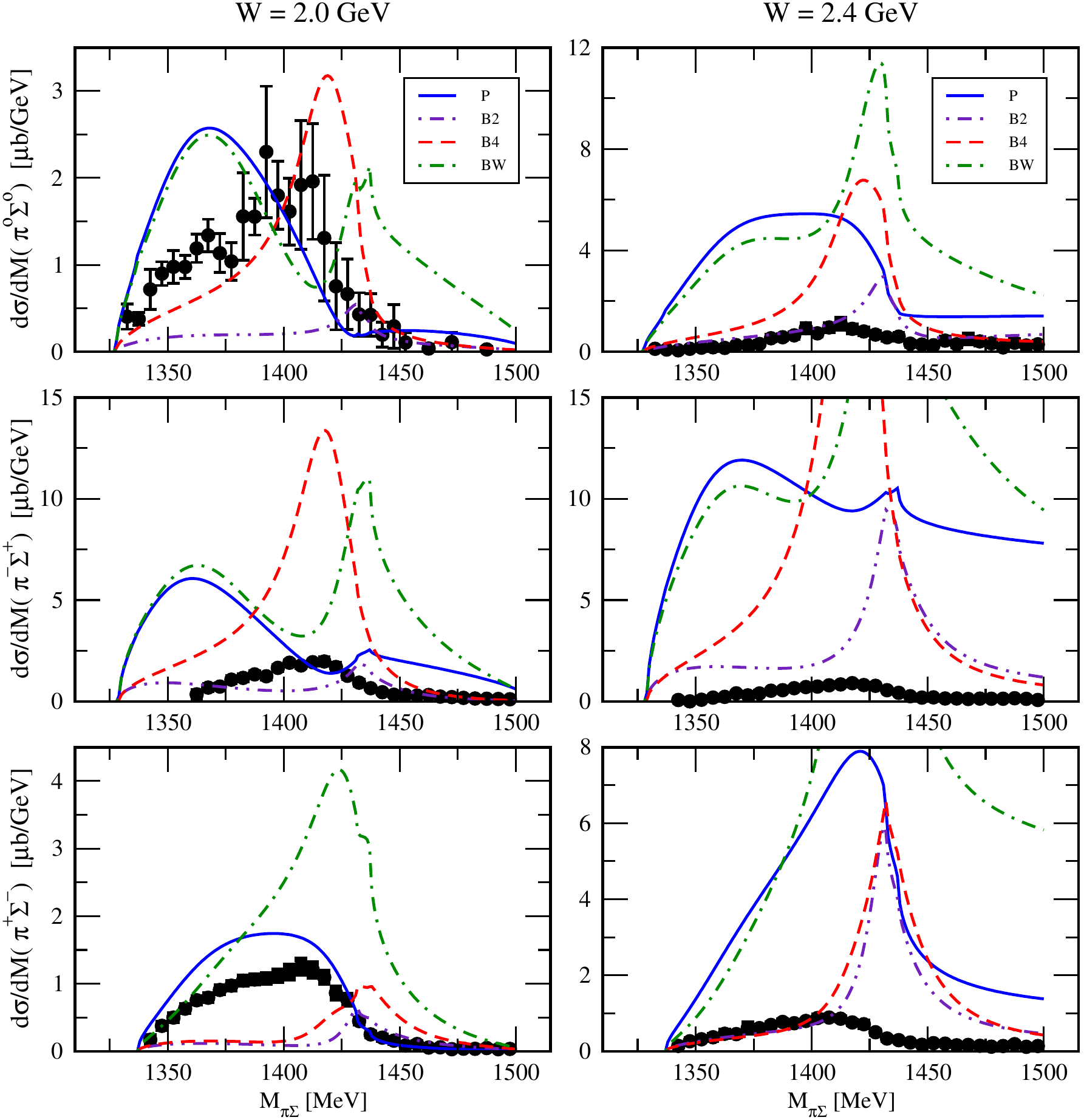}
    \caption{Comparison of $\pi\Sigma$ mass distributions calculated at two c.m.~energies, $W = 2.0$ and $2.4~{\rm GeV}$, 
    while employing $MB$ amplitudes generated by four different coupled-channel models tagged as P \cite{Bruns:2021krp}, 
    B2, B4 \cite{Mai:2014xna} and BW \cite{Sadasivan:2018jig}. The experimental data are taken from \cite{Moriya:2013eb}.}
    \label{fig:piSigDis}
\end{figure*}

It is also well known that the chirally motivated $\bar{K}N$ models provide very different predictions 
for the energies below the $\bar{K}N$ threshold as well as in the isovector sector \cite{Cieply:2016jby}. 
Thus, we felt it necessary to check the sensitivity of the calculated $\pi\Sigma$ mass spectra to variations of the $MB$ re-scattering amplitudes. In Fig.~\ref{fig:piSigDis} we show our results obtained at two sample energies, $W = 2.0$ (left figures) and $2.4~{\rm GeV}$ (right figures), employing four different $\bar{K}N$ models that provide the final state 
rescattering amplitudes $f_{0+}$ required in Eq.~(\ref{eq:modelA}). Besides the already mentioned Prague amplitudes \cite{Bruns:2021krp}, 
used to calculate the spectra shown in Fig.~\ref{fig:FSimpact} and tagged as P model here, we also show results 
obtained with three versions of the Bonn model amplitudes: B2, B4 \cite{Mai:2014xna}, and BW \cite{Sadasivan:2018jig}. 
For $W = 2.0~{\rm GeV}$, the continuous blue lines generated by the P model are exactly the same as those shown in Fig.~\ref{fig:FSimpact}.

%%%%%%%
%%%%%%%
%%%%%%%
%%%%%%%

The first impression one gets from Fig.~\ref{fig:piSigDis} is that, taking aside the $\pi^{0}\Sigma^{0}$ and $\pi^{+}\Sigma^{-}$ 
results at $W = 2.0~{\rm GeV}$, the theoretical predictions provide much higher cross sections than those observed experimentally. 
We remark that similarly large cross sections of $d\sigma/dM_{\pi\Sigma} \approx 5-10\;\mu{\rm b/GeV}$ at the peak mass 
were predicted in the early Refs.~\cite{Nacher:1998mi, Lutz:2004sg} as well. As we will illustrate below, however, there is 
a hope that the magnitude of the $\pi\Sigma$ mass spectra can be reduced by tuning the rescattering contribution to the 
photoproduction amplitude. We also note that our formalism is based on LO ChPT which is expected to work much better at low 
energies. Specifically, a kaon momentum of $|\vec{q}_{K}|(W=2\,\mathrm{GeV}, M_{\pi\Sigma}=1.35\,\mathrm{GeV}) \approx 350\,\mathrm{MeV}$ 
is still reasonable for three-flavor ChPT. However, at $W=2.4~{\rm GeV}$ the kaon momentum is much larger, $|\vec{q}_{K}| \approx 717$ MeV, 
probably too large for the LO ChPT photo-kernel. Therefore, the worse description of the observed mass distributions 
at $W=2.4~{\rm GeV}$ can be anticipated. 

\begin{figure*}[t]
    \centering
    \includegraphics[width=0.8\textwidth]{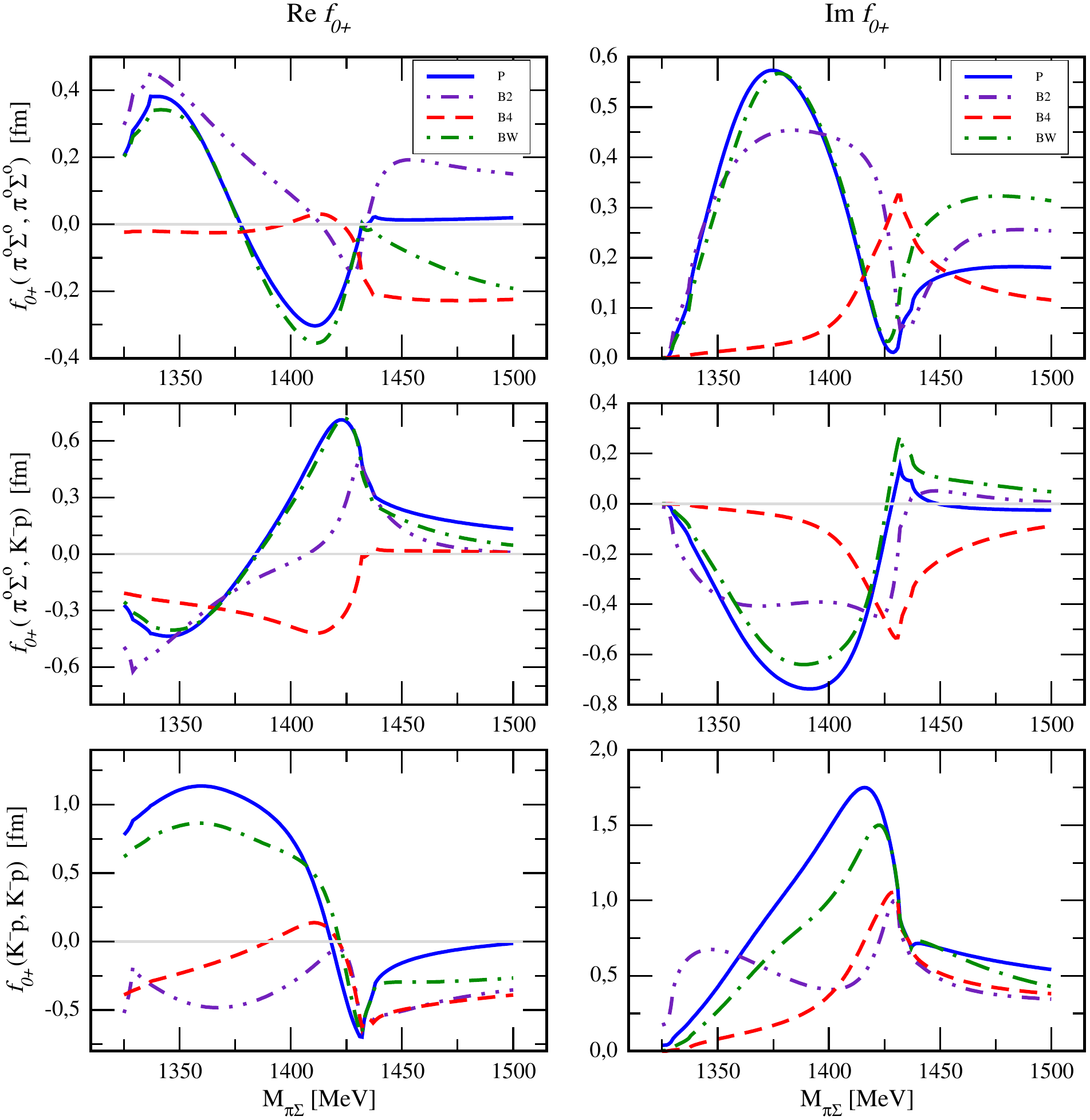}
    \caption{The real (left) and imaginary (right) parts of s-wave scattering amplitudes for reactions 
        $\pi^{0}\Sigma^{0} \rightarrow \pi^{0}\Sigma^{0}$, $K^{-}p \rightarrow \pi^{0}\Sigma^{0}$, and $K^{-}p \rightarrow K^{-}p$. The amplitudes were generated by four different $\bar{K}N$ coupled-channel models specified in the text. The P model amplitude of the inelastic $K^{-}p \rightarrow \pi^{0}\Sigma^{0}$ process is shown with an opposite sign for an easier comparison with the other models.}
    \label{fig:MBampl}
\end{figure*}

It is interesting to note the close proximity of the theoretical predictions made with the P and BW model amplitudes. Despite the Prague and Bonn approaches to $\pi\Sigma - \bar{K}N$ coupled-channel interactions do differ significantly in several aspects including the sets of experimental data their model parameters were fitted to, the P and BW models generate quite similar $\pi\Sigma$ photoproduction mass spectra up to about $1400$ MeV. This can be understood by comparing the respective rescattering amplitudes which we demonstrate in Fig.~\ref{fig:MBampl}. There, one can see that both models generate very similar $MB$ amplitudes for energies up to the $\bar{K}N$ threshold.

%%%%%%%
%%%%%%%
%%%%%%%
%%%%%%%

The other two Bonn models, B2 and B4, were in fact used in fits that included the $\pi\Sigma$ photoproduction data from the CLAS experiment~\cite{Mai:2014xna}. However, in those fits a simplified phenomenological approach was adopted to model the photoproduction part (see also Ref.~\cite{Roca:2013av}) of the process in terms of makeshift energy-dependent constants used to multiply the amplitudes responsible for the $MB$ rescattering in the final state. When compared with our present formulation of the photoproduction amplitude given in Eq.~(\ref{eq:modelA}) the B2 and B4 models correspond to setting the tree graphs in the first term on the r.h.s. to zero and (at the same time) replacing them by energy-dependent constants in the second term. It is obvious that such an ad-hoc treatment cannot reflect fully the complexity of the photoproduction process and the resulting $MB$ amplitudes may not be quite reliable. However, the B2 and B4 amplitudes still describe the data well for $K^{-}p$ reactions and represent a good option to test the model dependence of our theoretical 
predictions. Looking at Fig.~\ref{fig:piSigDis} it seems that the available $\bar{K}N$ models are flexible enough to generate a peak in the $\pi\Sigma$ mass spectra at varied energies, and at least in the $\pi^{-}\Sigma^{+}$ 
channel provide even two peaks in case the data would support such option. The BW model predicts two peaks also in the $\pi^{0}\Sigma^{0}$ mass spectra, the one at higher mass quite close to the $\bar{K}N$ threshold where the chirally motivated approaches predict a resonance interpreted as a   $K^{-}p$ molecular state.

\begin{figure*}[t]
%   \centering
    \includegraphics[width=0.9\textwidth]{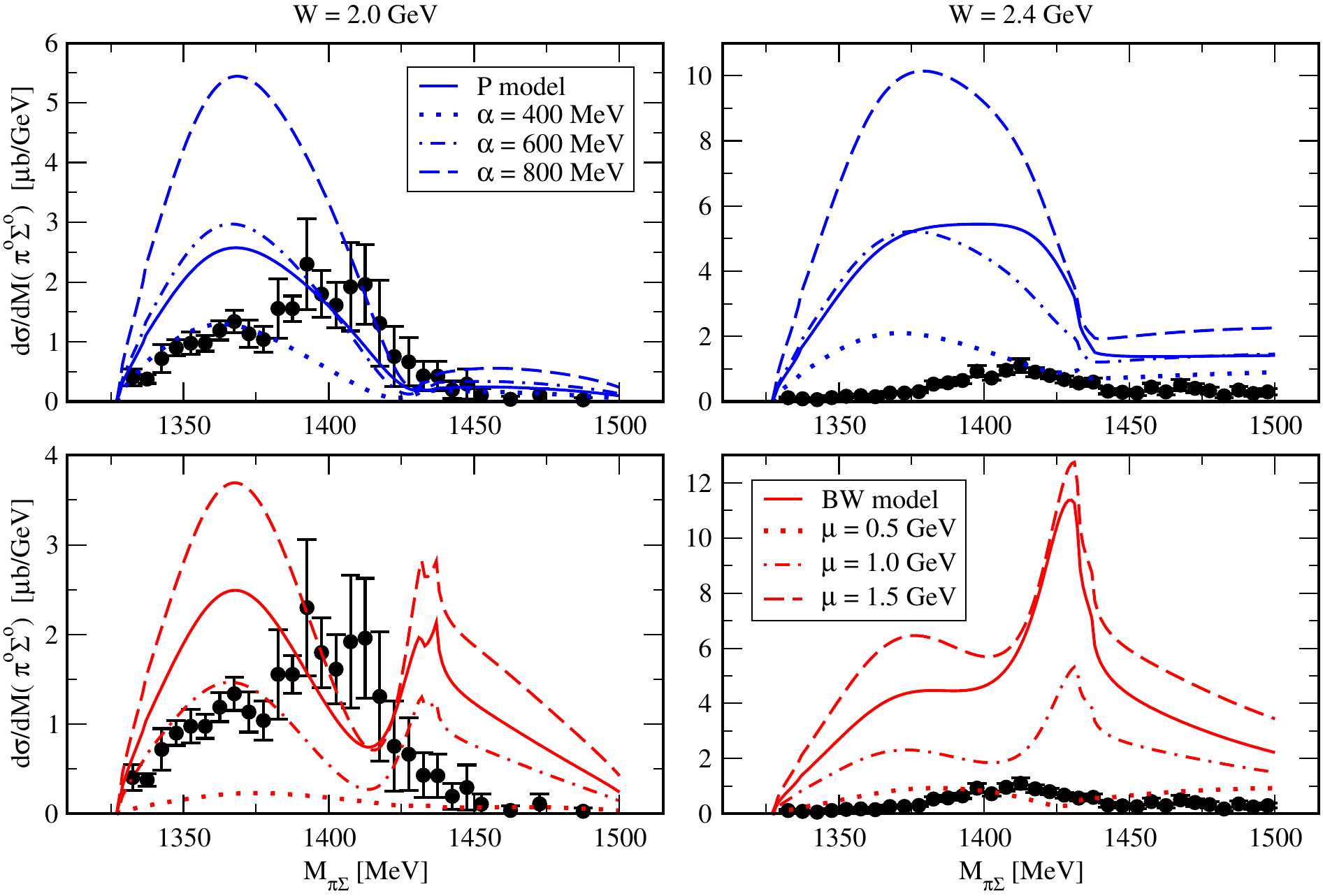}
    \caption{A demonstration of the impact caused by modifying the regularization scales adopted in the loop 
    functions connecting the tree level photoproduction graphs with the meson-baryon rescattering amplitudes. 
    The top (bottom) figures present the $\pi^{0}\Sigma^{0}$ mass spectra generated with the modified P model
    (BW model), the results obtained with the unmodified loop functions are shown for comparison as well.
    }
    \label{fig:FSmod}
\end{figure*}

Coming back to the the $\pi\Sigma$ mass spectra, we recall that generating the peak structures is only possible including 
the $MB$ rescattering in the final state as shown in \cref{fig:FSimpact}. Algebraically, this
is represented by the second term in Eq.~(\ref{eq:modelA}) and one may consider modifications of either the rescattering amplitude or of the loop function that connects it to the tree level photoproduction graphs. The first option would require a completely new fit to the experimental data including both, the $K^{-}p$ reactions data (at threshold and higher energies) as well as the $\pi\Sigma$ mass spectra discussed here. 
The complexity of such fits goes well beyond the scope of our current work, but we wish to illustrate how the magnitude of the photoproduction cross sections can be tuned by modifying the loop function $G(M_{\pi\Sigma})$ that enters Eq.~(\ref{eq:modelA}). This is demonstrated in Fig.~\ref{fig:FSmod} where we present the results obtained 
with different choices of the regularization scales -- the inverse range $\alpha$ for the P model and the mass scale $\mu$ 
in case of the BW model. For simplicity, the same scale $\alpha$ or $\mu$ is used for all ten $MB$ channels. The original mass spectra generated by the P and BW models (employing different regularization scales in different channels, matching those used to generate the rescattering amplitudes) are presented in the figure for comparison as well.

%%%%%%%
%%%%%%%
%%%%%%%
%%%%%%%

The calculated spectra shown in Fig.~\ref{fig:FSmod} illustrate that lower values of the regularization scales $\alpha$ or $\mu$ 
lead to smaller cross sections. It seems that for $\alpha \approx 400$ MeV or for $\mu \approx 1.0~{\rm GeV}$ the magnitude 
of the generated $\pi\Sigma$ distributions is comparable to the one found in the CLAS experiment. Of course, the peak position and structure of the generated mass spectra require additional modifications that are driven by the energy dependence of the adopted $f_{0+}$ amplitudes. As already stated, there is a room for such an adjustment since different $MB$ coupled-channel models provide varied energy dependence of the amplitudes in between the $\pi\Sigma$ and $\bar{K}N$ thresholds. 

Finally, we would like to comment on the feasibility of altering only the first $MB$ loop functions that appears in Eq.~(\ref{eq:modelA}) while keeping unchanged the rescattering amplitudes and the loop functions used in coupled-channel approaches that generate them. Since the vertices connecting the photoproduction tree graphs with the $MB$ rescattering 
part have a different structure than the $MB \rightarrow M'B'$ vertices one can consider altering the first, e.g., by imposing additional form factors on them, while keeping the latter intact. Thus, the adopted modifications of the first loop functions that we used to tune the magnitude of the $\pi\Sigma$ cross sections can effectively relate to modifications of the $\cal{M}$ vertex in Fig.~\ref{fig:MBFSI}. We note that a good reproduction of the CLAS data presented in Ref.~\cite{Nakamura:2015rta} required a combined impact of modifications of the first $MB$ loop functions, addition of phenomenological contact terms contributing to the $\cal{M}$ vertex as well as addition of form factors applied to it.

%%%%%%%%%%%%%%%%%%%%%%%%%%%%%%%%%%%%%%%%%%%%%%%%%%%%%%%%%%%%%%%%%%%%%%%%%%%%%%%%%%%%%%%%%%%%%%%
\section{Summary and Conclusion}
\label{sec:conclusion}
%%%%%%%%%%%%%%%%%%%%%%%%%%%%%%%%%%%%%%%%%%%%%%%%%%%%%%%%%%%%%%%%%%%%%%%%%%%%%%%%%%%%%%%%%%%%%%%

In this article, we  have outlined a methodology to incorporate the meson-baryon final-state interaction 
into the two-meson photoproduction amplitude through partial-wave amplitudes $f_{\ell\pm}$ constructed within 
the framework of \emph{Unitarized ChPT}. This formalism 
can be used to implement coupled-channel unitarity, low-energy theorems from ChPT and gauge invariance in the description 
of the photoproduction process. Our results obtained with two versions of modern $\bar{K}N$ models, adopted to generate 
the $MB$ amplitudes, demonstrate the crucial role played by the final-state interaction in the $\pi\Sigma$ photoproduction process. When the $MB$ rescattering is omitted, only the considered tree graphs 
contribute to the photoproduction amplitude, and the generated cross sections are rather small and flat. The peak structure observed in the mass spectra reported by the CLAS collaboration and related to a formation of the $\Lambda(1405)$ resonance appears only when the $MB$ rescattering in the final state is accounted for in our formalism. 

Our results show that the formalism is capable of reproducing the experimental data at c.m.~energies $W \approx 2.0~{\rm GeV}$ 
where the pertinent kaon momenta are relatively small, complying with the limitations of three-flavor ChPT. The situation 
is worse at higher energies where our predictions provide too large cross sections when compared with the CLAS data.
However, we would like to emphasize once again that these predictions are made without introducing any adjustment 
to the $\pi\Sigma - \bar{K}N$ coupled-channel models that were fitted to describe the data on $K^{-}p$ reactions, 
at a completely different sector of kinematics and energies. 
We have also refrained from introducing any additional mechanisms in the tree level photoproduction graphs 
that would go beyond the standard ChPT approach, nor applied any ad-hoc energy dependent factors to moderate 
the generated mass spectra. In this sense our current predictions are completely parameter-free. 
A much better agreement with the experimental data can be achieved by a combination of including the $\pi\Sigma$ 
photoproduction data in the fits of the $\bar{K}N$ models and maybe by applying an additional form factor 
to the $\mathcal{M}$ vertex depicted in Fig.~\ref{fig:MBFSI}. Moreover, we have shown that adopting different 
$\bar{K}N$ models for the $f_{0+}$ amplitudes leads to varied structure of the computed $\pi\Sigma$ mass distributions 
that can accommodate spectra with either one or two peaks. At the same time the magnitude of the calculated cross sections 
can be tuned by modifying the first $MB$ loop function or the $\mathcal{M}$ vertex.

Of course, it is also natural to ask about a role played by other kinds of final-state interaction inherent 
in the process, as, e.g., the pion-kaon interaction, irreducible three-body interactions, or even triangle-graph mechanisms 
as studied in Ref.~\cite{Wang:2016dtb}. In this respect we note that effects due to an enhanced final-state interaction 
in the other channels have already been subtracted in the CLAS data with which we compare our predictions \cite{Moriya:2013eb}. 
There, it was also reported that these effects have only a moderate impact on the $\pi\Sigma$ invariant 
mass spectra. Thus, it seems that a direct application of the approach proposed here is a reasonable strategy. In the future, one should also make efforts for a more complete description of the photoproduction process, 
along the lines explained so far. This would enable us to compare our predictions with a more complete subset of the provided data, as, e.g., the kaon angular distributions, which are presumably sensitive to final-state 
interaction in sectors not considered in this work.

%%%%%%%%%%%%%%%%%%%%%%%%%%%%%%%%%%%%%%%%%%%%%%%%%%%%%%%
\begin{center}
{\bf Acknowledgements}
\end{center}
The work of P.C.B. and A.C. was supported by the Grant Agency of the Czech Republic under Grant No.~19-19640S. 
M.M. acknowledges support by the Deutsche
Forschungsgemeinschaft (DFG, German Research Foundation) and the NSFC through the funds provided to the Sino-German
Collaborative Research Center CRC 110 “Symmetries
and the Emergence of Structure in QCD” (DFG Project-ID 196253076 -
TRR 110, NSFC Grant No.~12070131001).
%%%%%%%%%%%%%%%%%%%%%%%%%%%%%%%%%%%%%%%%%%%%%%%%%%%%%%%

%%%%%%%%%%%%%%%%%%%%%%%%%%%%%%%%%%%%%%%%%%%%
%%%%%%%%%%%%%%%%%%%%%%%%%%%%%%%%%%%%%%%%%%%%
\bibliography{BIB3}
%%%%%%%%%%%%%%%%%%%%%%%%%%%%%%%%%%%%%%%%%%%%
%%%%%%%%%%%%%%%%%%%%%%%%%%%%%%%%%%%%%%%%%%%%

\widetext
\begin{appendix}

\section{Projection on \texorpdfstring{$\mathbf{\pi\Sigma}$}{} partial waves}
\label{app:cc_formalism}
\def\theequation{\Alph{section}.\arabic{equation}}
\setcounter{equation}{0}

Here we provide explicit expressions for the s-wave projections of the photoproduction amplitude for $MB=\pi\Sigma$, 
but it is obvious that analogous expressions for the other meson-baryon-channels can be found by simple replacements 
of the appropriate masses. As a first step, we define the combinations
\begin{align}
  &\mathcal{C}_{0+}^{1} = \mathcal{M}_{1}' + (\sqrt{s}+m_{N})\mathcal{M}_{5}'  + (\sqrt{s}-M_{\pi\Sigma})\left(\mathcal{M}_{9}'+(\sqrt{s}+m_{N})\mathcal{M}_{13}'\right)\,,\\
  &\mathcal{C}_{0+}^{2} = \mathcal{M}_{1}' - (\sqrt{s}-m_{N})\mathcal{M}_{5}'  - (\sqrt{s}+M_{\pi\Sigma})\left(\mathcal{M}_{9}'-(\sqrt{s}-m_{N})\mathcal{M}_{13}'\right)\,,\\
  &\mathcal{C}_{0+}^{3} = \mathcal{M}_{1}' + (\sqrt{s}-M_{\pi\Sigma})\mathcal{M}_{9}'   
  + \frac{1}{2}(\sqrt{s}+m_{N})\bigl(\mathcal{M}_{2}'+(\sqrt{s}-m_{N})\mathcal{M}_{6}'  + (\sqrt{s}-M_{\pi\Sigma})\left(\mathcal{M}_{10}'+(\sqrt{s}-m_{N})\mathcal{M}_{14}'\right)\bigr)\,,\\
  &\mathcal{C}_{0+}^{4} = \mathcal{M}_{1}' - (\sqrt{s}+M_{\pi\Sigma})\mathcal{M}_{9}' 
  - \frac{1}{2}(\sqrt{s}-m_{N})\bigl(\mathcal{M}_{2}'-(\sqrt{s}+m_{N})\mathcal{M}_{6}'  - (\sqrt{s}+M_{\pi\Sigma})\left(\mathcal{M}_{10}'-(\sqrt{s}+m_{N})\mathcal{M}_{14}'\right)\bigr)\,,
\end{align}
employing the following abbreviations:
\begin{eqnarray}
  \mathcal{M}_{1}' &:=&  \overline{\mathcal{M}}_{1} - \frac{1}{3}(E_{\Sigma}^{\ast}-m_{\Sigma})\overline{\mathcal{M}}_{3}\,,\qquad 
  \mathcal{M}_{5}' :=  \overline{\mathcal{M}}_{5} + \frac{1}{3}(E_{\Sigma}^{\ast}-m_{\Sigma})\overline{\mathcal{M}}_{7}\,,\\
  \mathcal{M}_{9}' &:=&  \overline{\mathcal{M}}_{9} + \frac{1}{3}(E_{\Sigma}^{\ast}-m_{\Sigma})\overline{\mathcal{M}}_{11}\,,\qquad
  \mathcal{M}_{13}' :=  \overline{\mathcal{M}}_{13} - \frac{1}{3}(E_{\Sigma}^{\ast}-m_{\Sigma})\overline{\mathcal{M}}_{15}\,,\\
  \mathcal{M}_{2}' &:=&  \overline{\mathcal{M}}_{2} + \frac{1}{3M_{\pi\Sigma}}(4E_{\Sigma}^{\ast}-m_{\Sigma})\overline{\mathcal{M}}_{3}\,,\qquad
  \mathcal{M}_{6}' :=  \overline{\mathcal{M}}_{6} + \frac{1}{3M_{\pi\Sigma}}(4E_{\Sigma}^{\ast}-m_{\Sigma})\overline{\mathcal{M}}_{7}\,,\\
  \mathcal{M}_{10}' &:=&  \overline{\mathcal{M}}_{10} + \frac{1}{3M_{\pi\Sigma}}(4E_{\Sigma}^{\ast}-m_{\Sigma})\overline{\mathcal{M}}_{11}\,,\qquad
  \mathcal{M}_{14}' :=  \overline{\mathcal{M}}_{14} + \frac{1}{3M_{\pi\Sigma}}(4E_{\Sigma}^{\ast}-m_{\Sigma})\overline{\mathcal{M}}_{15}\,.
\end{eqnarray}
See Eq.~(\ref{eq:MmuDecomp}) for the definition of the $\mathcal{M}_{i}$, and Eq.~(\ref{eq:MbarApprox}) for the definition of the $\overline{\mathcal{M}}_{i}$. We point out that the structure functions $\mathcal{M}_{4,8,12,16}$ are always eliminated via the gauge-invariance constraints
\begin{eqnarray}
  (s-m_{N}^2)\mathcal{M}_{2}\, &\overset{!}{=}& \,(u_{\Sigma}-m_{\Sigma}^2)\mathcal{M}_{3} + (t_{K}-M_{K}^2)\mathcal{M}_{4}\,,\nonumber \\
  2\mathcal{M}_{1}+ (s-m_{N}^2)\mathcal{M}_{6}\, &\overset{!}{=}& \,(u_{\Sigma}-m_{\Sigma}^2)\mathcal{M}_{7} + (t_{K}-M_{K}^2)\mathcal{M}_{8}\,,\nonumber \\
  (s-m_{N}^2)\mathcal{M}_{10}\, &\overset{!}{=}& \,(u_{\Sigma}-m_{\Sigma}^2)\mathcal{M}_{11} + (t_{K}-M_{K}^2)\mathcal{M}_{12}\,,\nonumber \\
  2\mathcal{M}_{9}+ (s-m_{N}^2)\mathcal{M}_{14}\, &\overset{!}{=}& \,(u_{\Sigma}-m_{\Sigma}^2)\mathcal{M}_{15} + (t_{K}-M_{K}^2)\mathcal{M}_{16}\,.\label{eq:gaugeinv}
\end{eqnarray}
In the simple case of structure functions $\mathcal{M}_{i}$ independent of $t_{\Sigma},\,u_{\Sigma}$, vanishing for $i=3,7,11,15$, we have $\mathcal{M}_{i}'=\mathcal{M}_{i}$\,. In fact, we can easily construct a gauge-invariant amplitude $\mathcal{M}^{\mu}$ of such a simplified form, which yields a set of prescribed $\mathcal{C}_{0+}^{i}(s,M_{\pi\Sigma}^2,t_{K})$,
\begin{eqnarray}
  \mathcal{M}_{1}^{\mathcal{C}} &=& \frac{1}{4\sqrt{s}M_{\pi\Sigma}}\biggl((\sqrt{s}+M_{\pi\Sigma})(\sqrt{s}-m_{N})\mathcal{C}_{0+}^{1}   -(\sqrt{s}-M_{\pi\Sigma})(\sqrt{s}+m_{N})\mathcal{C}_{0+}^{2}\biggr)\,,\\
  \mathcal{M}_{2}^{\mathcal{C}} &=& \frac{1}{2\sqrt{s}M_{\pi\Sigma}}\biggl((\sqrt{s}+M_{\pi\Sigma})\mathcal{C}_{0+}^{3} + (\sqrt{s}-M_{\pi\Sigma})\mathcal{C}_{0+}^{4}\biggr)\,,\\
  \mathcal{M}_{3}^{\mathcal{C}} &=& 0\,,\\
  \mathcal{M}_{5}^{\mathcal{C}} &=& \frac{1}{4\sqrt{s}M_{\pi\Sigma}}\biggl((\sqrt{s}+M_{\pi\Sigma})\mathcal{C}_{0+}^{1} + (\sqrt{s}-M_{\pi\Sigma})\mathcal{C}_{0+}^{2}\biggr)\,,\\
  \mathcal{M}_{6}^{\mathcal{C}} &=& \frac{1}{2\sqrt{s}M_{\pi\Sigma}}\biggl(\frac{\sqrt{s}+M_{\pi\Sigma}}{\sqrt{s}+m_{N}}\left(\mathcal{C}_{0+}^{3}-\mathcal{C}_{0+}^{1}\right)  + \frac{\sqrt{s}-M_{\pi\Sigma}}{\sqrt{s}-m_{N}}\left(\mathcal{C}_{0+}^{2}-\mathcal{C}_{0+}^{4}\right)\biggr)\,,\\
  \mathcal{M}_{7}^{\mathcal{C}} &=&0\,,\\
  \mathcal{M}_{9}^{\mathcal{C}} &=& -\frac{1}{4\sqrt{s}M_{\pi\Sigma}}\biggl((\sqrt{s}-m_{N})\mathcal{C}_{0+}^{1} + (\sqrt{s}+m_{N})\mathcal{C}_{0+}^{2}\biggr)\,,\\
  \mathcal{M}_{10}^{\mathcal{C}} &=& \frac{1}{2\sqrt{s}M_{\pi\Sigma}}\left(\mathcal{C}_{0+}^{4}-\mathcal{C}_{0+}^{3}\right)\,,\\
  \mathcal{M}_{11}^{\mathcal{C}}&=&0\,,\\
  \mathcal{M}_{13}^{\mathcal{C}} &=& \frac{1}{4\sqrt{s}M_{\pi\Sigma}}\left(\mathcal{C}_{0+}^{2}-\mathcal{C}_{0+}^{1} \right)\,,\\
  \mathcal{M}_{14}^{\mathcal{C}} &=& \frac{1}{2\sqrt{s}M_{\pi\Sigma}}\left(\frac{\mathcal{C}_{0+}^{1}-\mathcal{C}_{0+}^{3}}{\sqrt{s}+m_{N}} + \frac{\mathcal{C}_{0+}^{2} - \mathcal{C}_{0+}^{4}}{\sqrt{s}-m_{N}}\right)\,,\\ \mathcal{M}_{15}^{\mathcal{C}}&=&0\,,
\end{eqnarray}
with $\mathcal{M}_{4,8,12,16}^{\mathcal{C}}$ accordingly fixed from the gauge-invariance constraints. Finally, we define
\begin{eqnarray*}
  \mathcal{A}_{0+}^{1} &:=& \sqrt{E_{\Sigma}^{\ast}+m_{\Sigma}}\sqrt{E_{\pi\Sigma}+M_{\pi\Sigma}}\,\left(\mathcal{C}_{0+}^{1}\right)\sqrt{E_{N}-m_{N}}\,/\sqrt{2M_{\pi\Sigma}}\,,\\
  \mathcal{A}_{0+}^{2} &:=& \sqrt{E_{\Sigma}^{\ast}+m_{\Sigma}}\sqrt{E_{\pi\Sigma}-M_{\pi\Sigma}}\,\left(\mathcal{C}_{0+}^{2}\right)\sqrt{E_{N}+m_{N}}\,/\sqrt{2M_{\pi\Sigma}}\,,\\
  \mathcal{A}_{0+}^{3} &:=& \sqrt{E_{\Sigma}^{\ast}+m_{\Sigma}}\sqrt{E_{\pi\Sigma}+M_{\pi\Sigma}}\,\left(\mathcal{C}_{0+}^{3}\right)\sqrt{E_{N}-m_{N}}\,/\sqrt{2M_{\pi\Sigma}}\,,\\
  \mathcal{A}_{0+}^{4} &:=& \sqrt{E_{\Sigma}^{\ast}+m_{\Sigma}}\sqrt{E_{\pi\Sigma}-M_{\pi\Sigma}}\,\left(\mathcal{C}_{0+}^{4}\right)\sqrt{E_{N}+m_{N}}\,/\sqrt{2M_{\pi\Sigma}}\,,
\end{eqnarray*}
where $E_{\pi\Sigma}$ is the c.m. energy of the $\pi\Sigma$ pair, $E_{\pi\Sigma}=\sqrt{s}-E_{K}=(s+M_{\pi\Sigma}^2-M_{K}^2)/(2\sqrt{s})$. We also note that $|\vec{q}_{K}|=\sqrt{E_{\pi\Sigma}^2-M_{\pi\Sigma}^2}\,$ and $E_{N}=(s+m_{N}^2)/(2\sqrt{s})$\,, 
$E_{\Sigma}^{\,\ast} = (M_{\pi\Sigma}^2+m_{\Sigma}^2-M_{\pi}^2)/(2M_{\pi\Sigma})$\,.

\end{appendix}

\end{document}